\documentclass[runningheads]{llncs}

\usepackage{color}
\usepackage{amssymb}

\usepackage{graphicx}
\usepackage{tablefootnote}
\begin{document}
\title{Noninvasive Estimation of Mean Pulmonary Artery Pressure Using MRI, Computer Models, and Machine Learning}
\titlerunning{Noninvasive Estimation of mPAP from MRI, Computer Models, and ML}

\author{Michal K. Grzeszczyk\inst{1}\orcidID{0000-0002-5304-1020} \and
Tadeusz Satława\inst{1} \and Angela Lungu\inst{2}\orcidID{0000-0002-4531-2791} \and
Andrew Swift\inst{3} \and Andrew Narracott\inst{3,4}\orcidID{0000-0002-3068-6192} \and Rod Hose \inst{3} \and Tomasz Trzcinski\inst{5,6,7}\orcidID{0000-0002-1486-8906} \and Arkadiusz Sitek\inst{1}\orcidID{0000-0002-0677-4002}}
\authorrunning{M. Grzeszczyk et al.}

\institute{$^1$ Sano Centre for Computational Medicine, Cracow, Poland \\
$^2$ Technical University of Cluj-Napoca, Cluj-Napoca, Romania \\
$^3$ The University of Sheffield, Sheffield, United Kingdom \\
$^4$ Insigneo Institute for in Silico Medicine, University of Sheffield, Sheffield, UK \\
$^5$ Warsaw University of Technology $^6$ Tooploox $^7$ Jagiellonian University of Cracow}

\maketitle
\begin{abstract}
Pulmonary Hypertension (PH) is a severe disease characterized by an elevated pulmonary artery pressure. The gold standard for PH diagnosis is measurement of mean Pulmonary Artery Pressure (mPAP) during an invasive Right Heart Catheterization. In this paper, we investigate noninvasive approach to PH detection utilizing Magnetic Resonance Imaging, Computer Models and Machine Learning. We show using the ablation study, that physics-informed feature engineering based on models of blood circulation increases the performance of Gradient Boosting Decision Trees-based algorithms for classification of PH and regression of values of mPAP. We compare results of regression (with thresholding of estimated mPAP) and classification and demonstrate that metrics achieved in both experiments are comparable. The predicted mPAP values are more informative to the physicians than the probability of PH returned by classification models. They provide the intuitive explanation of the outcome of the machine learning model (clinicians are accustomed to the mPAP metric, contrary to the PH probability). 

\end{abstract}
\section{Introduction}
Pulmonary Hypertension is a severe disease difficult to diagnose with multiple possible root causes \cite{MHoeper2017}. For many years, PH was identified if a mean Pulmonary Artery Pressure (mPAP) of a patient at rest was equal to or above 25 mmHg. Recently, it has been suggested to lower the threshold to 20 mmHg \cite{Simonneau2019}. The precise measurement of mPAP is non-trivial and requires conducting an invasive Right Heart Catheterization (RHC) – the gold standard for diagnosing PH. This procedure carries risks, requires patient’s preparation, trained staff, highly specialized equipment, it is expensive and time consuming. To lower the probability of complications it has to be performed at a specialized facility \cite{HoeperRHC}. 

Non-invasive estimation of mPAP using medical imaging, mathematical modeling, and machine learning (ML) is an option to avoid issues related with RHC. Mathematical models, such as a Windkessel model, allow diagnosis of the vascular system parameters \cite{Ae2008}. Different ML algorithms enable extracting knowledge about data samples and their performance usually increases with the addition of features from multiple domains.

In this paper, we present methods based on Gradient Boosting Decision Trees (GBDT) for non-invasive PH diagnosis. We use classic GBDT, DART (Dropouts meet Multiple Additive Regression Trees) \cite{vinayak2015dart} - a method utilizing dropouts of random trees during training - and GOSS (Gradient-based One-Side Sampling) \cite{ke2017lightgbm} – a technique that uses different than GBDT process of training (retaining samples with large gradients and randomly dropping the ones with low gradients). We conduct analysis on data from 352-patient cohort and perform two tasks: classification of PH and regression of mPAP. As predictors, we use demographics features, measurements derived from Magnetic Resonance Imaging (MRI) and features obtained from 0D and 1D mathematical models \cite{Lungu2014}. 

Our main contribution is the demonstration of the ablation study, which shows, that physics-informed feature engineering based on mathematical models of blood circulation increases the performance of ML algorithms for classification and regression of PH and values of mPAP, respectively. 
Another significant contribution of this paper is comparison of utilities of classification and regression approaches for the detection of PH. While the regression achieves similar classification metrics (after thresholding of estimated mPAP), the values of predicted mPAP are more informative to the physicians than the probability of PH returned by classification models. As such, they provide the intuitive explanation of the outcome of the machine learning model (clinicians are accustomed to the mPAP metric, contrary to the PH probability).

\section{Related work}

Multiple ML algorithms (utilizing features from various modalities like echocardiography, Computed Tomography (CT), or MRI) have been integrated for the purpose of the PH classification. In \cite{Leha2019}, five ML models were used and compared with each other. Boosted Classification Trees, Lasso Penalized Logistic Regression (LPLR), Random Forest (RF) for Regression, RF for Classification and Support Vector Machines (SVM) were adopted for mPAP prediction or PH classification basing on the echocardiographic measurements and basic patients characteristics (age, sex, BMI, body surface area). In \cite{Zhu2020}, echocardiographic data was used to distinguish between pre- and post-capillary PH with one of the nine tested ML models (SVM, AdaBoost, LR, RF, Decision Trees (DT), K-Nearest Neighbours, GBDT, LogitBoost and Linear Discriminant Analysis (LDA)). 

In \cite{Huang2020}, measurements derived from CT were used to train six ML classifiers to evaluate the probability of mPAP higher than 15 mmHg. Another approach was to record the heart sounds with a digital stethoscope to gather parameters for PH classification using LDA \cite{Elgendi2018}. The analysis of the sounds revealed specific patterns in PH patients. In \cite{Dennis2010}, it was noted that the sounds collected by phonocardiogram can be applied for binary classification of PH with SVM. In \cite{Lungu2016}, it was shown that MRI measurements combined with parameters from 0D and 1D computational models can be successfully used for PH and non-PH patients classification with DT. In our approach, we study the impact of mathematical models parameters on classification and regression. We also show the comparable performance of PH diagnosis with GBDT-based models in both tasks.

With the rise of Deep Learning (DL), multiple approaches of detecting PH directly from images, videos, or electrocardiography (ECG) signals were investigated. For example, chest X-Ray images can be utilized for binary classification of potential PH patients using Capsule Network with residual blocks \cite{Kusunose2020}. In \cite{Zou2020}, three popular DL networks (ResNet50, Xception and Inception V3) were trained as predictors of PH. As shown in \cite{Kwon2020}, an ensemble neural network can pose as a screening tool for PH from a 12-lead ECG signal.

ML can also be utilized for determining patients at risk of having Pulmonary Arterial Hypertension (PAH) from clinical records. In \cite{Kiely2019}, it was shown that GBDT can help in screening for PAH based on their medical history.  ML-based tools were also developed for the purpose of blood pressure estimation - in \cite{Zhang2019}, Support Vector Machine Regression (SVR) models were applied for the prediction of the patient’s blood pressure from the physiological data. Another example is an application of Multilayer Perceptron (MLP) for regression of systolic blood pressure using basic knowledge about patients (BMI, age, habits etc.) \cite{Wu2014}.

\section{Methods}

In this section, we describe our approaches to noninvasive PH diagnosis. We present the details of our dataset and introduce mathematical models which enabled the acquisition of physics-informed features. Finally, we train GBDT-based models on multiple feature sets to perform mPAP regression and PH classification experiments.

\subsection{PH dataset}

\begin{table}

\caption{PH dataset with patient related data, parameters derived from 0D and 1D models and measurements from MRI imaging. In the appendix (section~\ref{sec_appendix}) we provide explanations for the feature names. P-value tests a null hypothesis that the coefficient of the univariate linear regression between a feature and mPAP is equal to zero.}\label{tab_dataset}
\begin{tabular}{|l | c c c | c c c| c |}
\cline{2-7}
\multicolumn{1}{l|}{} &  \multicolumn{3}{l|}{No PH} & \multicolumn{3}{|l|}{PH}\\ \cline{3-3}
\hline
\textbf{feature}	&	\textbf{cnt} &	\textbf{mean} &	 \textbf{std}	&	\textbf{cnt} &	\textbf{mean} &	\textbf{std} & \textbf{p-value}\\
\hline
mPAP, mmHg	&	66	&	19.67	&	3.34 &		286 &	46.95 &	13.08 & \\
\hline
\textbf{Demographics} &&&&&&&\\
\hline
age, years		& 66	&	56.61	&	13.78	&	286 &	61.69 &	14.24 & 0.242\\
gender, female/male	&	66 &	43/23 &	&		286 &	173/113 & & 0.549\\
who, no.		& 56	&	2.52	&	0.54 &		285 &	3.04 &	0.44 & $<0.001$ \\
bsa, $m^2$		& 65	&	1.88	&	0.25 &	 	286 &	1.82 &	 0.22 & 0.24\\
\hline
\textbf{0D and 1D models} &&&&&&&\\
\hline
\textit{R$_{d}$}, $kg/m^4s$		& 66	&	6.08E+07	&	4.94E+07	&	286 &	1.46E+08 &	2.53E+08 & $<0.001$\\
\textit{R$_{c}$}, $kg/m^4s$		& 66	&	7.94E+06	&	7.80E+06 &		286 &	9.17E+06 &	1.87E+07 & 0.072\\
\textit{C}, $m^4s^2/kg$		& 66	&	9.92E-09	&	6.71E-09	&	284 &	3.94E-04 &	6.65E-03 & 0.669\\
\textit{R$_{tot}$}, $kg/m^4s$	&	66	&	6.83E+07	&	5.38E+07 &	286 &	1.56E+08 &	2.62E+08 & $<0.001$\\
\textit{W$_{b}$/W$_{tot}$}	&	66	&	0.24	&	0.10 &		286 &	0.39	& 0.11 & $<0.001$\\
\hline
\textbf{MRI} &&&&&&&\\
\hline
rac\_fiesta, \%	&	66	&	26.39	&	15.43 &		286 &	13.68 &	8.93 & $<0.001$\\
syst\_area\_fiesta, $cm^2$	&	66	&	7.62	&	2.17 &		286 &	9.78 &	2.78 & $<0.001$\\
diast\_area\_fiesta, $cm^2$	&	66	&	6.08	&	1.71 &		286 &	8.66 &	2.57 & $<0.001$\\
rvedv, $mL$ &		66	&	118.93	&	36.00 &		286 &	159.58 &	58.27 & $<0.001$\\
rvedv\_index, $mL/m^2$ &		66	&	53.78	&	21.83 &		286 &	73.92 &	39.39 & $<0.001$\\
rvesv, $mL$ &		66	&	55.41	&	20.68 &		286 &	102.48 &	49.92 & $<0.001$\\
rvesv\_index, $mL/m^2$ &		66	&	24.64	&	10.84 &		286 &	47.63 &	30.19 & $<0.001$\\
rvef, \% &		66	&	53.32	&	9.86 &		286 &	38.05 &	13.59 & $<0.001$\\
rvsv, $mL$ &		66	&	63.52	&	22.61 &		286 &	57.15 &	23.39 & 0.026\\
rvsv\_index, $mL/m^2$	&	66	&	29.14	&	13.90 &		286 &	26.32 &	15.02 & 0.292\\
lvedv, $mL$ &	66	&	116.57	&	33.09 &		286 &	91.30 &	27.33 & $<0.001$\\
lvedv\_index, $mL/m^2$	&	66	&	53.16	&	21.90 &		286 &	41.25 &	19.20 & $<0.001$ \\
lvesv, $mL$  &		66	&	34.27	&	15.66 &		286 &	31.32 &	14.56 & 0.23\\
lvesv\_index, $mL/m^2$	&	66	&	16.85	&	16.81 &		286 &	14.01 &	8.18 & 0.194\\
lvef, \%	& 	66	&	71.13	&	8.54 &		286 &	65.81 &	10.92 & $<0.001$\\
lvsv, $mL$	&	66	&	82.30	&	23.30 &	 	286 &	59.97 &	19.93 & $<0.001$\\
lvsv\_index, $mL/m^2$	&	66	&	38.07	&	16.20	&	286 &	27.20 &	13.51 & $<0.001$\\
rv\_dia\_mass, $g$	&	66	&	22.62	&	6.80 &		283 &	44.48 &	25.47 & $<0.001$\\
lv\_dia\_mass, $g$	&	66	&	91.47	&	27.71 &		286 &	90.64 &	24.98 & 0.436\\
lv\_syst\_mass, $g$ &		66	&	111.74	&	32.17 &		286 &	99.83 &	26.39 & $<0.001$\\
rv\_mass\_index, $g/m^2$ &		66	&	10.44	&	4.94 &		285 &	20.94 &	15.09 & $<0.001$\\
lv\_mass\_index, $g/m^2$	&	59	&	40.90	&	17.87 &		243 &	39.84 &	18.99 & 0.442\\
sept\_angle\_syst, degrees &		66	&	139.95	&	11.68 &		286 &	172.51 &	22.11 & $<0.001$ \\
sept\_angle\_diast, degrees &		66	&	134.21	&	8.28 &		286 &	145.01 &	11.93 & $<0.001$\\
4ch\_la\_area, $mm^2$	&	66	&	1921.95	&	387.56 &		286 &	1785.95 &	556.53 & $<0.001$\\
4ch\_la\_length, $mm^2$	& 	66	&	55.76	&	7.86 &		286 &	55.62 &	8.60 & 0.412\\
2ch\_la\_area, $mm^2$	&	66	&	1764.62	&	496.75 &		286 &	1901.67 &	545.35 & 0.855\\
2ch\_la\_length, $mm^2$	&	66	&	48.66	&	9.08 &		286 &	52.12 &	9.33 & 0.166\\
la\_volume, $mL$	&	66	&	55.22	&	17.96 &		286 &	54.16	 & 25.36 & 0.005\\
la\_volume\_index, $mL/m^2$ &		66	&	24.95	&	10.14 &		286 &	23.24 &	10.45 & 0.042\\
ao\_qflowpos, $L/min$ &		65	&	6.09	&	1.50	 &	285 &	5.29 &	1.50 & $<0.001$\\
ao\_qfp\_ind, $L/min/m^2$ &		65	&	2.79	&	1.18 &		285	 & 2.44 &	1.15 & 0.003\\
pa\_qflowpos, $L/min$ &		66	&	5.50	&	1.84	&	284	& 5.00 &	1.97 & 0.006\\
pa\_qflowneg, $L/min$ &		66	&	0.62	&	0.59	&	285	& 1.07 &	0.83 & $<0.001$\\
pa\_qfn\_ind, $L/min/m^2$ &		66	&	9.70	&	7.19	&	284 & 	17.49 &	9.85 & $<0.001$\\
systolic\_area\_pc, $mm^2$ &		66	&	731.05	&	236.42	&	284 &	950.17 &	268.98 & $<0.001$\\
diastolic\_area\_pc, $mm^2$	&	66	&	619.82	&	162.71	&	284 &	866.42 &	244.57 & $<0.001$\\
rac\_pc, \%	&	66	&	17.02	&	13.70 &		284	& 10.01 &	8.14 & $<0.001$\\

\hline
\end{tabular}
\end{table}

Table \ref{tab_dataset} presents the available features of patients who were suspected with PH and underwent MRI and RHC within 48 hours.

The medical procedures were performed at the Sheffield Pulmonary Vascular Disease Unit. The RHC procedure was conducted with a balloon-tipped 7.5-Fr thermodilution catheter.  The PH was defined if measured mPAP $\geq$ 25 mmHg. Using these criteria out of the cohort of 352 patients 286 were diagnosed with PH. From 286 patients with PH, 142 had Pulmonary Arterial Hypertension, 86 had Chronic Thromboembolic PH, 35 PH cases were due to lung diseases (e.g. Chronic Obstructive Pulmonary Disease), 15 cases were associated with left heart disease. The cause of PH in the rest of patients was either multifactorial or unknown. All of the available data samples are part of the ASPIRE Registry (Assessing the Severity of Pulmonary Hypertension In a Pulmonary Hypertension REferral Centre) \cite{Hurdman2012}.

MRI images were captured with 1.5-tesla whole-body scanner (GE HDx, GE Healthcare, Milwaukee) with an 8-channel cardiac coil. The images were acquired in the supine position during a breath hold. The balanced steady state free precession (bSSFP) sequences were spatially and temporally synchronized with the 2D phase contrast (PC) images of the Main Pulmonary Artery (MPA) using cardiac gating. Short-axis and four-chamber cardiac images were also collected. The features from MRI were obtained as in \cite{swift2012diagnostic}. \textit{A(t)} area of the MPA was extracted from the semi-automatically segmented bSSFP images. The blood flow through MPA (\textit{Q(t)}) was extracted from the segmented areas overlaid on PC images. Using those measurements 0D- and 1D-model features were derived.

To prepare the feature dataset for the training of ML models we fill the missing values using linear interpolation. 
We encode categorical features to numerical values and scale all the features to have means of 0 and variances of 1.

\subsection{Features derived from models of blood circulation}
The cardiovascular system (CVS) is a closed circuit with the main purpose of transporting oxygenated blood to organs and tissues \cite{Quarteroni2022}. It comprises especially from heart, blood and vessels. One of the main components of the CVS is the pulmonary circulation. The target of the pulmonary circulation is to transport the deoxygenated blood from the right ventricle through MPA and other arteries to lungs and deliver the oxygenated blood to the left ventricle \cite{Jain2021}. Since CVS can be described by its haemodynamics and structure of heart and vessels, the computational models based on the simplified representation of CVS were introduced \cite{Shi2011}. Those models range from 0D models simulating the global haemodynamics (e.g. resistance and compliance of the system) to 3D models representing the complex behaviour of vessels and the blood flow over time. In \cite{Lungu2014}, two models (0D and 1D) based on MRI measurements for the diagnosis of PH were proposed.

\subsubsection{0D model.}
0D models are often based on the hydraulic-electrical analogue - the blood flow and electrical circuits have many computational similarities \cite{Shi2011}. For example, the friction in the vessel can be identified as resistance R, the blood pressure as voltage and the flow-rate as current. Thus, by applying electrical laws (e.g. Kirchhoff's law, Ohm's law), the simplified representation of the CVS can be achieved. 0D modelling of CVS started with the implementation of the two-element Windkessel model \cite{Ae2008}. Different variants of this model appeared in the literature and it was applied to simulate pulmonary circulation \cite{Grant1987,Slife1990}. The 3-element (\textit{R$_{c}$CR$_{d}$}) Windkessel model comprises of the capacitor \textit{C} characterizing the compliance of the pulmonary circulation and two resistors \textit{R$_{c}$} and \textit{R$_{d}$} representing the resistance proximal and distal to the capacitor respectively.

In \cite{Lungu2014}, \textit{R$_{c}$CR$_{d}$} model was applied to capture the characteristics of PH and non-PH patients. In this model, the sum of two resistors can be interpreted as the ratio between mean pressure and mean flow (pulmonary vascular resistance - PVR) and \textit{C} indicates the compliance of the pulmonary arteries. To optimize the parameters of 0D model for the specific patient, two MRI imaging techniques of MPA were used: PC and bSSFP. The bSSFP images were segmented to find the area of MPA (\textit{A(t)}) over time. Then, the segmented regions were overlayed over PC images to capture the blood flow through MPA (\textit{Q(t)}). Having the \textit{Q(t)} and pressure \textit{p(t)} (derived from the measured MPA radius) the parameters of the Windkessel model which were best describing the relationship between \textit{Q(t)} and \textit{p(t)} over time could be derived.

\subsubsection{1D model.}
The simplified representation of the pulmonary vasculature is multiple elastic tubes with numerous branches. 1D models often analyse the propagation of the pressure and flow waves in such structures. The 1D equations of the waves travelling through elastic tubes are derived from Navier-Stokes equations. In \cite{Lungu2014}, the analysis of the power of the pressure waves was performed. The pressure wave was broken down into forward and backward-travelling elements (since vessels are rugged and twisted, some waves are bouncing off the vessel walls and travel backward). It was assumed and confirmed that the power of the backward wave in relation to the total wave power was greatly higher in PH cases than in healthy ones. As diseased pulmonary vasculature contains more deposits and stenoses the ratio of the backward wave power to the total wave power (represented as \textit{W$_{b}$/W$_{tot}$}) is higher than in the healthy one.

\subsection{Machine Learning for PH detection}

\subsubsection{mPAP regression.}

The decision whether the patient is suffering from PH is more important to the doctors than the actual value of mPAP. However, the non-invasive prediction of the PH occurrence together with the predicted value of mPAP is more informative to the clinicians. Therefore, we decide to conduct two experiments: mPAP regression and PH classification. 

To find the best ML algorithm for mPAP regression we train three models based on GBDT: classic GBDT, DART and GOSS. We use mPAP feature as the ground truth for our models. We find the best hyper parameters for the models using Bayes optimization with 8-fold cross-validation (CV). We optimize them for 200 iterations with minimizing Mean Squared Error (MSE) as the optimization target. Then, using the best found parameters we train the models with leave one out cross validation (LOOCV) and MSE as the objective function. We measure MSE, Root MSE (RMSE) and Mean Absolute Error (MAE) as regression metrics of the model. 

We assume that mPAP $\geq$ 25 mmHg is a positive PH diagnosis. With this assumption, we compute the binary classification metrics after thresholding the predicted and measured mPAP with 25 value. We calculate accuracy, sensitivity, specificity, True Positives (TP), False Positives (FP), True Negatives (TN), False Negatives (FN). To compare the impact of different feature sets on the results (demographics, MRI, mathematical models), we repeat the procedure of hyper parameter optimization, models training with LOOCV and metrics collection for different combinations of features. We compare results of all the approaches. 

Additionally, we train four other than boosted tree ML models on all features and compare the metrics with GBDT-based methods using LOOCV-derived metrics. These additional methods are MLP, SVR, AdaBoost and RF. 

\subsubsection{PH classification.}

We conduct the binary PH classification, similarly to mPAP regression. We binarize mPAP feature with 25 mmHg threshold and train three GBDT-based models on different variations of feature sets, previously optimizing the hyper parameters using Bayes optimization. The optimization is handled for 200 iterations with 8-fold stratified CV to ensure similar distribution of positive and negative samples over each fold. The optimization goal is the maximum area under the receiver operating characteristic (ROC) curve. 

We train GBDT, DART and GOSS on best found parameters with LOOCV. We calculate binary classification metrics: area under ROC curve (AUC), sensitivity, specificity, accuracy, TP, FP, TN and FN. To compute the binary classification metrics we use multiple thresholding strategies: youden - maximization of specificity + sensitivity, f1 - maximization of f1 metric (harmonic mean of precision and recall), closest01 - the point which is closest to (0,1) point on the ROC curve, concordance - maximization of the product of sensitivity and specificity. 

\section{Results}
In this section, we present results of our experiments. We analyze, through the ablation study, the impact of different feature sets on models performance and compare metrics achieved by regression and classification models. In our case, the ablation study means the removal of feature sets before the training to understand their contribution to the overall performance of ML models. We also show, that regression models can be utilized as a tool for PH classification.

\subsection{mPAP regression}

\begin{table}[t!]
\caption{Results of mPAP value regression with LOOCV. Models trained on demographics, MRI-derived features and 0D and 1D models parameters. P-value is calculated based on MAE against DART model.}
\label{tab_regression_results}
    \centering
    \begin{tabular}{|c|c|c|c|c|c|c|c|c|}
    \hline
        Method & MAE & RMSE & MSE & $R^2$ & sensitivity & specificity & accuracy & p-value\\ \hline
        MLP & 7.71 & 10.37 & 107.50 & 0.58 & 0.93 & 0.55 & 0.86 & $<0.001$\\ \hline
        SVR & 7.29 & 9.39 & 88.14 & 0.65 & 0.95 & 0.55 & 0.88 & $<0.001$\\ \hline
        AdaBoost & 6.92 & 8.92 & 79.59 & 0.69 & 0.97 & 0.41 & 0.87 & $<0.001$\\ \hline
        RandomForest & 6.55 & 8.64 & 74.59 & 0.71 & 0.95 & 0.56 & 0.88 & 0.003\\ \hline
        GOSS & 6.44 & 8.38 & 70.22 & 0.72 & 0.96 & 0.67 & 0.90 & $<0.001$ \\ \hline
        GBDT & 5.95 & 7.91 & 62.55 & 0.75 & 0.96 & 0.74 & 0.92 & 0.93 \\ \hline
        DART & 5.94 & 7.85 & 61.66 & 0.76 & 0.95 & 0.74 & 0.91 &  \multicolumn{1}{|l}{}\\ \cline{1-8}
    \end{tabular}
\end{table}

\begin{table}
    \caption{Ablation study over the combinations of available feature sets (demographics, MRI, 0D and 1D models). P-value is calculated against models trained on all features.}
    \centering
    \begin{tabular}{|l|l|l|l|l|l|l|l|}
    \hline
        demographics & \checkmark & ~ & ~ & \checkmark & ~ & \checkmark & \checkmark \\ \hline
        0D and 1D models & ~ & \checkmark & ~ & \checkmark & \checkmark & ~ & \checkmark \\ \hline
        MRI & ~ & ~ & \checkmark & ~ & \checkmark & \checkmark & \checkmark \\ \hline \hline
        regression (MAE) & \multicolumn{7}{|l|}{} \\ \hline
        GOSS & 11.09 & 9.16 & 6.93 & 8.44 & 6.77 & 6.51 & 6.44 \\
        \quad p-value & $<0.001$ & $<0.001$ & 0.007 & $<0.001$ & 0.012 & 0.645 &  \\ \hline
        GBDT & 10.85 & 9.14 & 6.69 & 8.33 & 6.49 & 6.34 & 5.95 \\ 
        \quad p-value & $<0.001$ & $<0.001$ & $<0.001$ & $<0.001$ & $<0.001$ & 0.012 &  \\ \hline
        DART & 11.01 & 9.35 & 6.76 & 8.43 & 6.20 & 6.20 & 5.94 \\
        \quad p-value & $<0.001$ & $<0.001$ & $<0.001$ & $<0.001$ & 0.058 & 0.083 &  \\ \hline \hline
        classification (AUC) & \multicolumn{7}{|l|}{} \\ \hline
        GOSS & 0.74 & 0.87 & 0.91 & 0.89 & 0.95 & 0.93 & 0.95 \\ 
        \quad p-value & $<0.001$ & $<0.001$ & $<0.001$ & $<0.001$ & 0.117 & $<0.001$ &  \\ \hline
        GBDT & 0.77 & 0.85 & 0.93 & 0.88 & 0.94 & 0.93 & 0.94 \\
        \quad p-value & $<0.001$ & $<0.001$ & 0.593 & $<0.001$ & 0.147 & 0.017 &  \\ \hline
        DART & 0.79 & 0.85 & 0.93 & 0.88 & 0.95 & 0.93 & 0.95 \\ 
        \quad p-value & $<0.001$ & $<0.001$ & 0.005 & $<0.001$ & 0.028 & $<0.001$ &  \\ \hline
    \end{tabular}

    \label{tab_ablation}
\end{table}

Table \ref{tab_regression_results} presents results of regression experiments. The lowest regression metrics are achieved by DART MAE=5.94, RMSE=7.85 and MSE=61.66. GBDT has marginally better classification metrics with sensitivity=0.96 (DART, 0.95), specificity=0.74 (DART, 0.74) and accuracy=0.92 (DART, 0.91). The difference between DART and GBDT results is not statistically significant (p-value=0.93). Additionally, GBDT-based methods outperform other tested ML algorithms: RF, AdaBoost, SVR and MLP with RF achieving the lowest MAE (6.55) out of all compared methods (p-value=0.003). 

Table \ref{tab_ablation} shows results of the ablation study. For all models, MAE drops with different combinations of feature sets (demographics, mathematical models, MRI) as opposed to only one feature set. The lowest MAE when a single feature set is used is achieved for MRI-derived measurements (GOSS, 6.93; GDBT 6.69; DART, 6.76). However, the combination of all available feature sets yields the best performance (GOSS, 6.44; GDBT 5.95; DART, 5.94). The physics-informed feature engineering performed by the addition of 0D and 1D models parameters improves metrics obtained in the regression.

\begin{figure}

\centering
\includegraphics[width=0.7\textwidth]{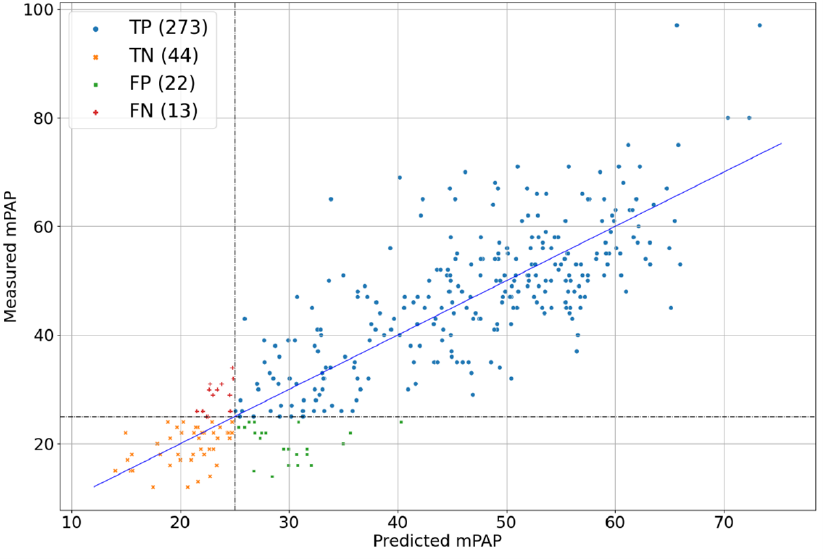}
\includegraphics[width=0.7\textwidth]{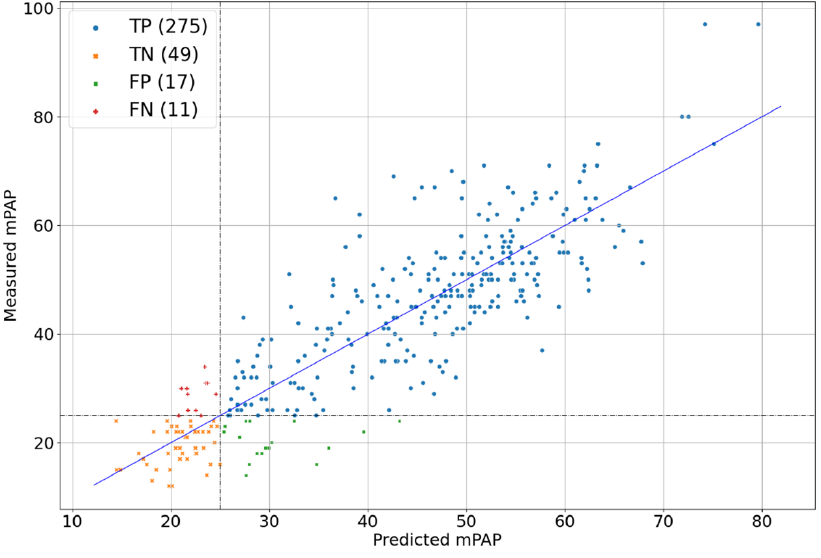}
\includegraphics[width=0.7\textwidth]{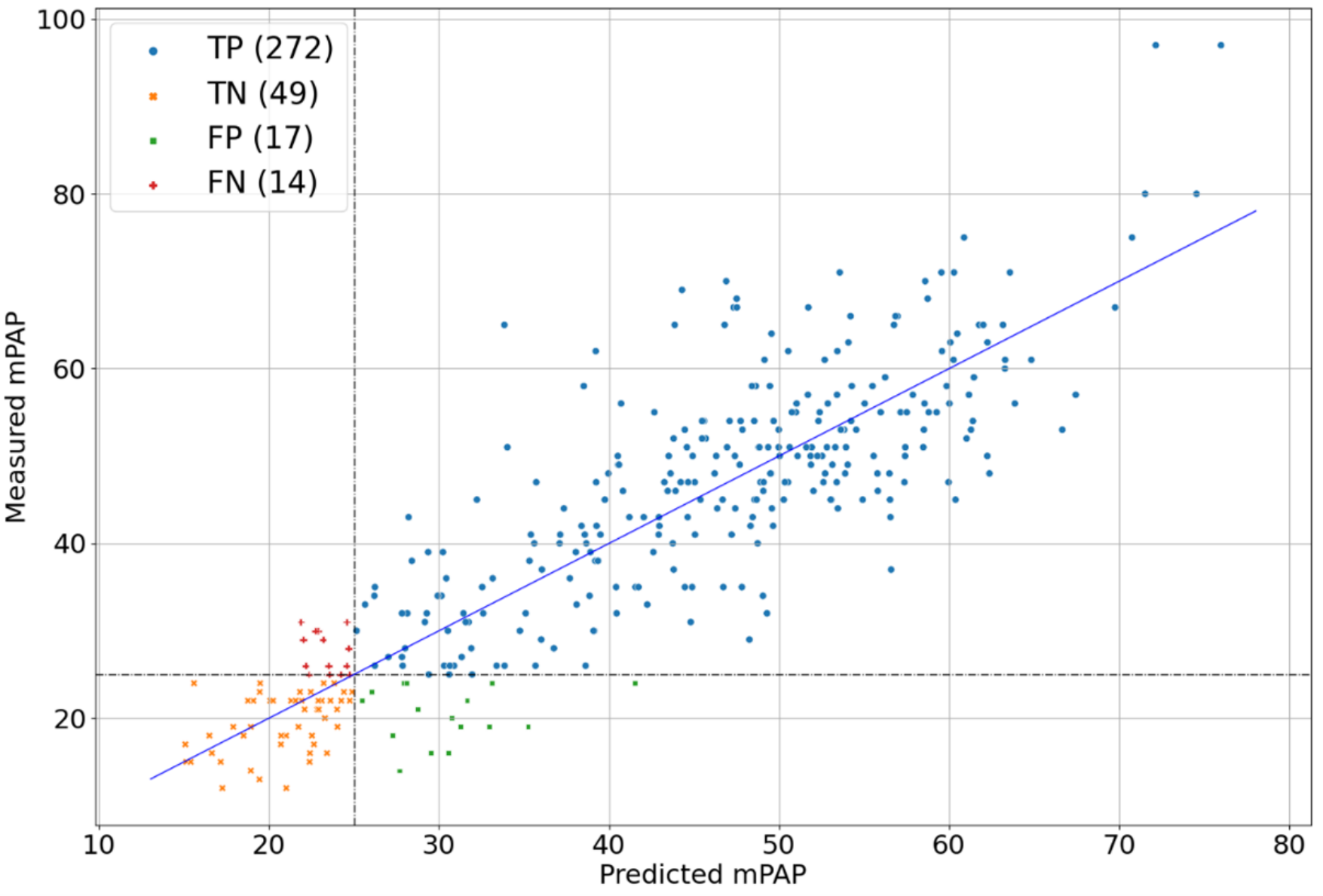}

\caption{Measured vs predicted values of mPAP with $\geq 25$ mmHg thresholding for models trained on all parameters: GOSS (top, $R^2$=0.72, p-value $<0.001$), GBDT (middle, $R^2$=0.75, p-value $<0.001$), DART (bottom, $R^2$=0.76, p-value $<0.001$).} 
\label{fig_regression_m_vs_p}
\end{figure}

The relations between predicted and measured mPAP values are shown in Figure \ref{fig_regression_m_vs_p}. The addition of mathematical models features decreases the number of FP and FN (calculated with 25 mmHg threshold) even though the models were trained on the MSE which is a regression objective function. Only 17 predictions are FP and 11 are FN for GBDT (DART; 17 FP, 14 FN). For GOSS, GBDT and DART, only one FP sample was predicted as having higher mPAP than 40 mmHg. The measured value for this patient during RHC was 24 mmHg which by current indicators means PH positive patient. In case of mPAP above 45 mmHg, all samples are predicted positively meaning high confidence of this model above that value. All false negative samples have the predicted values above 20 mmHg.

\subsection{PH classification}

\begin{figure}[t!]
\minipage{0.33\textwidth}
\includegraphics[width=\linewidth]{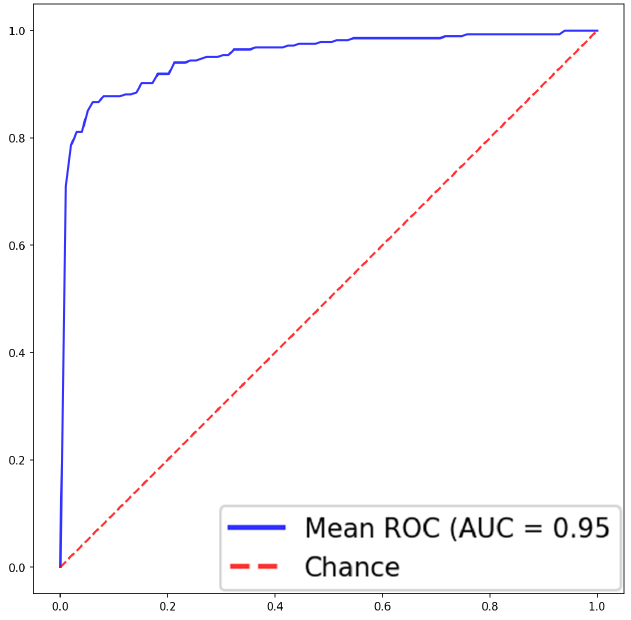}
\endminipage
\minipage{0.33\textwidth}
\includegraphics[width=\linewidth]{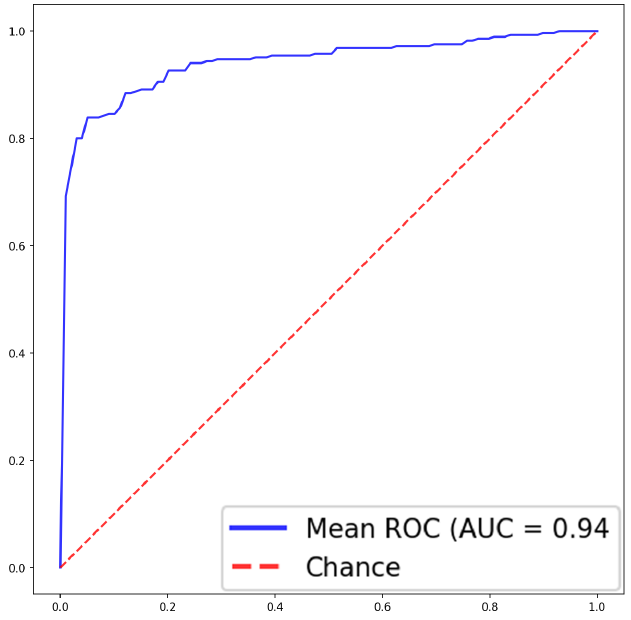}
\endminipage
\minipage{0.33\textwidth}
\includegraphics[width=\linewidth]{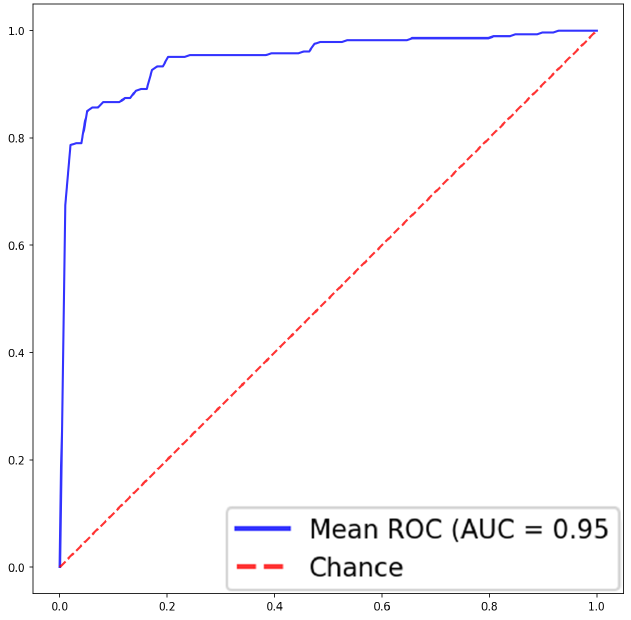}
\endminipage

\caption{ROC curves for GBDT-based classification models trained on all features: GOSS (left), GBDT (middle), DART (right).}
\label{fig_roc}
\end{figure}

In the PH classification experiments, the impact of 0D and 1D models parameters is also significant (Table~\ref{tab_ablation}). For the single feature set, the highest AUC is achieved for models trained on MRI-derived parameters (GOSS, 0.91; DART, 0.93; GDBT, 0.93). The addition of features from mathematical models improves the performance and acquires the same AUC as for the models trained on all parameters (GOSS, 0.95; GDBT 0.94; DART, 0.95). 
Table \ref{tab_classification_results} shows the detailed results of PH classification models trained on all features. The highest AUC is achieved for GOSS and DART models. Those models have the highest specificity (GOSS, 0.94; GBDT, 0.95; DART, 0.95) when thresholding their predicted probabilities with youden or concordance strategies. However, the classification of PH patients is a task in which we would like to detect as many positive patients as possible (maximizing sensitivity) while retaining reasonably high specificity (the percentage of accurately stating that no PH is present). Such an approach is most closely achieved with maximizing f1 metric as the thresholding strategy. With this strategy, DART predictions yield best metrics with sensitivity of 0.95, specificity of 0.8 and accuracy of 0.92. The results are comparable with the best regression metrics (sensitivity=0.96, specificity=0.74 and accuracy=0.92). The FN had mPAP close to 25 mmHg (with a maximum of 33 mmHg) and relatively small PVR, meaning, that no severe PH case was misclassified. Half of the FP had mPAP higher than 20 mmHg. The ROC curves for the three models are presented in Figure \ref{fig_roc}. 

\begin{table}[t!]
\caption{Results of PH classification with LOOCV. Models trained on demographics, MRI-derived features and 0D and 1D models parameters. Metrics sens (sensitivity), spec (specificity), acc (accuracy) are given for multiple thresholding strategies: youden, concordance, 01 (closest01), f1 (maximizing f1 metric).}
\label{tab_classification_results}
    \centering
    \begin{tabular}{|l||l||l|l|l||l|l|l||l|l|l||l|l|l|}
        \cline{3-14}
        \multicolumn{2}{l|}{} &  \multicolumn{3}{l||}{youden} & \multicolumn{3}{|l||}{concordance} & \multicolumn{3}{|l||}{01} & \multicolumn{3}{|l|}{f1}\\ \hline
        Method & AUC & sens & spec & acc & sens & spec & acc & sens & spec & acc & sens & spec & acc \\ \hline
        GOSS & 0.95 & 0.88 & 0.94 & 0.88 & 0.88 & 0.94 & 0.88 & 0.88 & 0.92 & 0.89 & 0.97 & 0.68 & 0.91 \\ \hline
        GBDT & 0.94 & 0.84 & 0.95 & 0.86 & 0.84 & 0.95 & 0.86 & 0.84 & 0.95 & 0.86 & 0.94 & 0.76 & 0.91 \\ \hline
        DART & 0.95 & 0.85 & 0.95 & 0.87 & 0.85 & 0.95 & 0.87 & 0.87 & 0.92 & 0.88 & 0.95 & 0.8 & 0.92 \\ \hline
    \end{tabular}
\end{table}

\section{Discussion}
The noninvasive assessment of mPAP is a difficult task. In a clinical setting the pressure is measured through the invasive RHC. The models presented in this paper enable the prediction of mPAP in a noninvasive way using information about patients, measurements derived from multiple MRI images and mathematical models. The combination of all features acquired from different domains brings the best results. The physics-informed feature engineering improves the assessment of mPAP. The modelling of MPA haemodynamics enables the quantification of physiological markers that enhance the quality of predictions. While MRI is not a widely used test in PH diagnosis, we showed that it can be utilized for an accurate, noninvasive mPAP estimation.

What is more, as our knowledge about the disease progresses, the thresholds and definitions of PH may change. Our regression models are not restricted to the 25 mmHg threshold set before the training. Depending on the current and future state of PH classification, the predicted mPAP can be interpreted in different ways. Classification models return the probability of patient having PH - the probability depends on the assumed threshold for PH. In this setting, the regression models are more flexible and can be used as additional information regarding the patient's state to help in determining the final diagnosis, even, if the definition of PH changes. In case the regression model is used for classification only, the predicted mPAP poses as an explanation of the diagnoses. As shown in Figure \ref{fig_regression_m_vs_p}, the confidence in a positive PH diagnosis can be stronger as the predicted mPAP gets higher. Above the predicted value of 45 mmHg all patients were diagnosed with PH. All the positive samples that were misclassified as negative have the predicted mPAP over 20 mmHg which can be considered elevated. In other words we have not observed any critical failures of our models.

Nevertheless, clinicians are mostly interested in the final diagnosis of the ML models. We show that classification models achieve similar metrics as the regression models: sensitivity=0.95, specificity=0.8 and accuracy=0.92 achieved by DART for classification, in comparison to sensitivity=0.96, specificity=0.74 and accuracy=0.92 achieved by GBDT for regression. It is important to notice that the impact of the models performance by the features from mathematical models is more clearly represented in the classification task, because the described mathematical models were created for discrimination between PH and non-PH patients \cite{Lungu2014}. Parameters derived from those models pose as an accurate PH/non-PH differentiation mechanism and prediction of mPAP from those parameters may be a harder task. However, the addition of features derived from 0D and 1D models improves the regression metrics as well.

\section{Conclusion}
In this paper, we investigated the impact of physics-informed feature engineering on the performance of GBDT-based models for mPAP regression and PH classification. We showed that parameters from 0D and 1D mathematical models improve the metrics of tested models. Comparison of the results revealed that the PH diagnosis may be performed by regression models achieving similar metrics as the classification models. The provided, predicted mPAP value increases the confidence in the final diagnosis. Future works may include improvements in the feature engineering, utilizing deep learning to predict mPAP directly from MRI images or testing our methods on external datasets.

\section{Acknowledgements}

This publication is partly supported by the European Union’s Horizon 2020 research and innovation programme under grant agreement Sano No. 857533 and the International Research Agendas programme of the Foundation for Polish Science, co-financed by the European Union under the European Regional Development Fund.
This research was partly funded by Foundation for Polish Science (grant no POIR.04.04.00-00-14DE/18-00 carried out within the Team-Net program co-financed by the European Union under the European Regional Development Fund), National Science Centre, Poland (grant no 2020/39/B/ST6/01511). The authors have applied a CC BY license to any Author Accepted Manuscript (AAM) version arising from this submission, in accordance with the grants’ open access conditions.

\bibliographystyle{splncs04}
\bibliography{bibliography}

\begin{thebibliography}{10}
\providecommand{\url}[1]{\texttt{#1}}
\providecommand{\urlprefix}{URL }
\providecommand{\doi}[1]{https://doi.org/#1}

\bibitem{Dennis2010}
Dennis, A., et~al.: Noninvasive diagnosis of pulmonary hypertension using heart
  sound analysis. Computers in Biology and Medicine  \textbf{40},  758--764 (9
  2010). \doi{10.1016/j.compbiomed.2010.07.003}

\bibitem{Elgendi2018}
Elgendi, M., Bobhate, P., Jain, S., Guo, L., Rutledge, J., Coe, Y., Zemp, R.,
  Schuurmans, D., Adatia, I.: The voice of the heart: Vowel-like sound in
  pulmonary artery hypertension. Diseases  \textbf{6} (2018).
  \doi{10.3390/diseases6020026}, \url{www.mdpi.com/journal/diseases}

\bibitem{galie2009guidelines}
Galie, N., et~al.: Guidelines for the diagnosis and treatment of pulmonary
  hypertension: the task force for the diagnosis and treatment of pulmonary
  hypertension of the european society of cardiology (esc) and the european
  respiratory society (ers), endorsed by the international society of heart and
  lung transplantation (ishlt). European heart journal  \textbf{30}(20),
  2493--2537 (2009)

\bibitem{Grant1987}
Grant, B.J., Paradowski, L.J.: Characterization of pulmonary arterial input
  impedance with lumped parameter models. American Journal of Physiology-Heart
  and Circulatory Physiology  \textbf{252},  H585--H593 (3 1987).
  \doi{10.1152/ajpheart.1987.252.3.H585}

\bibitem{HoeperRHC}
Hoeper, M.M., Lee, S.H., Voswinckel, R., et~al.: {Complications of right heart
  catheterization procedures in patients with pulmonary hypertension in
  experienced centers}. J Am Coll Cardiol  \textbf{48}(12),  2546--2552 (Dec
  2006)

\bibitem{MHoeper2017}
Hoeper, M.M., et~al.: Pulmonary hypertension. Dtsch Arztebl Int  \textbf{114},
  73--84 (2017). \doi{10.3238/arztebl.2017.0073}

\bibitem{Huang2020}
Huang, L., Li, J., Huang, M., Zhuang, J., Yuan, H., Jia, Q., Zeng, D., Que, L.,
  Xi, Y., Lin, J., Dong, Y.: Prediction of pulmonary pressure after glenn
  shunts by computed tomography-based machine learning models. European
  Radiology  \textbf{30},  1369--1377 (2020). \doi{10.1007/s00330-019-06502-3},
  \url{https://doi.org/10.1007/s00330-019-06502-3}

\bibitem{Hurdman2012}
Hurdman, J., Condliffe, R., Elliot, C., Davies, C., Hill, C., et~al.: Aspire
  registry: Assessing the spectrum of pulmonary hypertension identified at a
  referral centre. European Respiratory Journal  \textbf{39},  945--955 (4
  2012). \doi{10.1183/09031936.00078411}

\bibitem{Jain2021}
Jain, V., Bordes, S., Bhardwaj, A.: Physiology, Pulmonary Circulatory System.
  StatPearls Publishing (2021)

\bibitem{ke2017lightgbm}
Ke, G., et~al.: Lightgbm: A highly efficient gradient boosting decision tree.
  Advances in neural information processing systems  \textbf{30},  3146--3154
  (2017)

\bibitem{Kiely2019}
Kiely, D.G., et~al.: Utilising artificial intelligence to determine patients at
  risk of a rare disease: idiopathic pulmonary arterial hypertension. Pulmonary
  Circulation  \textbf{9} (10 2019). \doi{10.1177/2045894019890549}

\bibitem{Kusunose2020}
Kusunose, K., Hirata, Y., Tsuji, T., Kotoku, J., Sata, M.: Deep learning to
  predict elevated pulmonary artery pressure in patients with suspected
  pulmonary hypertension using standard chest x ray. Scientific Reports
  \textbf{10} (12 2020). \doi{10.1038/S41598-020-76359-W}

\bibitem{Kwon2020}
Kwon, J.M., Kim, K.H., Inojosa, J.M., Jeon, K.H., Park, J., Oh, B.H.:
  Artificial intelligence for early prediction of pulmonary hypertension using
  electrocardiography. The Journal of Heart and Lung Transplantation
  \textbf{39},  805--814 (8 2020). \doi{10.1016/j.healun.2020.04.009}

\bibitem{Leha2019}
Leha, A., Hellenkamp, K., Unsöld, B., Mushemi-Blake, S., Shah, A.M.,
  Hasenfuß, G., Seidler, T.: A machine learning approach for the prediction of
  pulmonary hypertension. PLoS ONE  \textbf{14} (10 2019).
  \doi{10.1371/journal.pone.0224453}

\bibitem{Lungu2014}
Lungu, A., Wild, J.M., Capener, D., Kiely, D.G., Swift, A.J., Hose, D.R.: Mri
  model-based non-invasive differential diagnosis in pulmonary hypertension.
  Journal of Biomechanics  \textbf{47},  2941--2947 (9 2014).
  \doi{10.1016/j.jbiomech.2014.07.024}

\bibitem{Lungu2016}
Lungu, A., Swift, A.J., Capener, D., Kiely, D., Hose, R., Wild, J.M.: Diagnosis
  of pulmonary hypertension from magnetic resonance imaging-based computational
  models and decision tree analysis. Pulmonary Circulation  \textbf{6},
  181--190 (6 2016). \doi{10.1086/686020}

\bibitem{Quarteroni2022}
Quarteroni, A., Manzoni, A., Vergara, C.: The cardiovascular system:
  Mathematical modelling, numerical algorithms and clinical applications. Acta
  Numerica  \textbf{26},  365--590 (2017). \doi{10.1017/S0962492917000046},
  \url{https://doi.org/10.1017/S0962492917000046}

\bibitem{Shi2011}
Shi, Y., Lawford, P., Hose, R.: Review of zero-d and 1-d models of blood flow
  in the cardiovascular system. BioMedical Engineering OnLine  \textbf{10}, ~33
  (12 2011). \doi{10.1186/1475-925X-10-33}

\bibitem{Simonneau2019}
Simonneau, G., et~al.: Haemodynamic definitions and updated clinical
  classification of pulmonary hypertension. European Respiratory Journal
  \textbf{53} (1 2019). \doi{10.1183/13993003.01913-2018}

\bibitem{Slife1990}
Slife, D.M., et~al.: Pulmonary arterial compliance at rest and exercise in
  normal humans. American Journal of Physiology-Heart and Circulatory
  Physiology  \textbf{258},  H1823--H1828 (6 1990).
  \doi{10.1152/ajpheart.1990.258.6.H1823}

\bibitem{swift2012diagnostic}
Swift, A.J., Rajaram, S., Condliffe, R., et~al.: Diagnostic accuracy of
  cardiovascular magnetic resonance imaging of right ventricular morphology and
  function in the assessment of suspected pulmonary hypertension results from
  the aspire registry. Journal of Cardiovascular Magnetic Resonance
  \textbf{14}(1),  1--10 (2012)

\bibitem{vinayak2015dart}
Vinayak, R.K., Gilad-Bachrach, R.: Dart: Dropouts meet multiple additive
  regression trees. In: Artificial Intelligence and Statistics. pp. 489--497.
  PMLR (2015)

\bibitem{Ae2008}
Westerhof, N., Lankhaar, J.W., Westerhof, B.E.: The arterial windkessel. Med
  Biol Eng Comput pp. 131--141 (2008). \doi{10.1007/s11517-008-0359-2}

\bibitem{Wu2014}
Wu, T.H., Pang, G.K.H., Kwong, E.W.Y.: Predicting systolic blood pressure using
  machine learning. 2014 7th International Conference on Information and
  Automation for Sustainability: "Sharpening the Future with Sustainable
  Technology", ICIAfS 2014  (3 2014). \doi{10.1109/ICIAFS.2014.7069529}

\bibitem{Zhang2019}
Zhang, B., Ren, H., Huang, G., Cheng, Y., Hu, C.: Predicting blood pressure
  from physiological index data using the svr algorithm. BMC Bioinformatics
  \textbf{20} (2 2019). \doi{10.1186/s12859-019-2667-y}

\bibitem{Zhu2020}
Zhu, F., Xu, D., Liu, Y., Lou, K., He, Z., et~al.: Machine learning for the
  diagnosis of pulmonary hypertension. Kardiologiya  \textbf{60},  96--101
  (2020). \doi{10.18087/cardio.2020.6.n953}

\bibitem{Zou2020}
Zou, X.L., et~al.: A promising approach for screening pulmonary hypertension
  based on frontal chest radiographs using deep learning: A retrospective
  study. PloS one  \textbf{15}(7) (2020). \doi{10.1371/journal.pone.0236378},
  \url{https://doi.org/10.1371/journal.pone.0236378}

\end{thebibliography}

\section{Appendix}
\label{sec_appendix}

Acronyms used in Table~\ref{tab_dataset} and their explanations: mPAP: mean pulmonary arterial pressure measured during RHC procedure, who: WHO functional PAH score \cite{galie2009guidelines}, bsa: body surface area, \textit{R$_{d}$}: distal resistance calculated from 0D model, \textit{R$_{c}$}: proximal resistance, \textit{C}: total pulmonary compliance, \textit{R$_{tot}$}: total resistance, \textit{W$_{b}$/W$_{tot}$}: backward pressure wave to the total wave power, rac\_fiesta: pulmonary arterial relative area change from bSSFP MRI, systolic\_area\_fiesta: syst area of MPA from bSSFP, diast\_area\_fiesta: diastolic area of MPA from bSSFP, rvedv: right ventricle end diastolic volume, rvedv\_index: rv end diastolic volume index, rvesv: rv end systolic volume, rvesv\_index: rv end systolic volume index, rvef: right ventricle ejection fraction, rvsv: rv stroke volume, rvsv\_index: rvsv index, lvedv: left ventricle end diastolic volume, lvedv\_index: lvedv index, lvesv: lv end systolic volume, lvesv\_index: lvesv index, lvef: lv ejection fraction, lvsv: lv stroke volume, lvsv\_index: lvsv index, rv\_dia\_mass: rv diastolic mass, lv\_dia\_mass: lv diastolic mass, lv\_syst\_mass: lv systolic mass, rv\_mass\_index: rv diastolic mass index, lv\_mass\_index: lv diastolic mass index, sept\_angle\_syst: systolic septal angle, sept\_angle\_diast: diastolic septal angle, 4ch\_la\_area: left atrium area 4 chamber, 4ch\_la\_length: la length 4 chamber, 2ch\_la\_area: left atrium area 2 chamber, 2ch\_la\_length: la length 2 chamber, la\_volume: la volume, la\_volume\_index: la volume index, ao\_qflowpos: aortic positive flow, ao\_qfp\_ind: aortic positive flow index, pa\_qflowpos: PA positive flow, pa\_qflowneg: PA negative flow, pa\_qfn\_ind: PA negative flow index, systolic\_area\_pc: systolic MPA area from PC, diastolic\_area\_pc: diastolic MPA area from PC, rac\_pc: relative area change of MPA from PC.

\end{document}